\documentstyle[12pt]{article}

\begin{document}

\newfam\msafam
\font\tenmsa=msam10        \textfont\msafam=\tenmsa
\font\sevenmsa=msam7       \scriptfont\msafam=\sevenmsa
\font\fivemsa=msam5        \scriptscriptfont\msafam=\fivemsa

\newfam\msbfam
\font\tenmsb=msbm10        \textfont\msbfam=\tenmsb
\font\sevenmsb=msbm7       \scriptfont\msbfam=\sevenmsb
\font\fivemsb=msbm5        \scriptscriptfont\msbfam=\fivemsb
\def\Bbb{\fam\msbfam \tenmsb}

\makeatletter                    
\@addtoreset{equation}{section}  
\makeatother                     
\renewcommand{\theequation}{\thesection.\arabic{equation}}

\title{Rigged Hilbert spaces and time-asymmetry: the case of the upside-down simple harmonic
oscillator}

\author{ Mario Castagnino*\ddag \dag \thanks{ Email: Castagni@iafe.uba.ar},
   Roberto Diener\ddag ,\\ Luis Lara\dag \ and
 Gabriel Puccini\dag \\
{\small  * IAFE (c.c. 67, suc. 28, C.P. 1428 Bs. As., Argentina)}
\\
{\small \ddag Dpto. de F\'{\i}sica, Universidad de Bs. As., Argentina}
\\
{\small \dag Facultad de Ciencias Exactas, Ingenier\'{\i}a y Agrimensura, Av.
 Pellegrini
250,}
\\
\small{2000 Rosario, Argentina}}

\maketitle

\begin{abstract}
The upside-down simple harmonic oscillator system is studied in the contexts of quantum
mechanics and classical statistical mechanics. It is shown that in order to study in a simple
manner the creation and decay of a physical system by ways of Gamow vectors we must
formulate the theory in  a time-asymmetric fashion, namely using two different rigged Hilbert
spaces to describe states evolving towards the past and the future. The spaces defined in the
contexts of quantum and classical statistical mechanics are shown to be directly
related by the Wigner function.
\end{abstract}

\vskip 2pc


\vskip 2pc
Number of manuscript pages: 36

\newpage

\openup2\jot

Proposed running head: RIGGED HILBERT SPACES ...

Proofs should be sent to: Mario Castagnino
			
IAFE
c.c. 67, suc. 28, C.P. 1428, Buenos Aires, Argentina

Email: Castagni@iafe.uba.ar

\section{Introduction}

In this work we will study the motion  of a particle subject to an upside-down simple harmonic oscillator potential 

\begin{equation}
V(q) = - {\frac{1 }{2}} m \omega^2 q^2  \label{potential}
\end{equation}

 in the contexts of  quantum mechanics and  classical statistical mechanics.
The aim of the paper is threefold:

1.-We solve the problem using a nonconventional technique. In fact,
 we find that the evolution of the system can be entirely expressed in terms
of idealized states that decay or grow exponentially (usually called Gamow vectors). In order to give mathematical
meaning to these states we are forced to work in the framework of a rigged
Hilbert space (RHS). In quantum mechanics the  Gamow vectors are then defined as generalized
eigenvectors of the Hamiltonian operator with complex eigenvalues, 
 being these  the poles of the scattering matrix when it is extended to
the complex plane.
The Gamow vectors can be used in `generalized spectral expansions' similar
to those found by the Gel'fand-Maurin theorem, but in order to use them
we have to define two different test function  spaces, that we shall call  $
\Phi _{+}$ and $\Phi _{-}$. A similar result is found in classical statistical mechanics. 

2.- We will prove that with this mathematical structure we have set the basis to introduce a time asymmetry
in the theory. Let us explain this further.
The vectors in  $\Phi _{+}$ can be expanded as linear combinations of the
decaying Gamow vectors while  the vectors in $\Phi _{-}$ can be expanded as
linear combinations of the growing  Gamow vectors. Since every physical
state (either quantum or statistical mechanical) must decay both towards the
future and the past, it is mathematically sound to represent the state of a
physical system when it evolves towards the future (i.e. from an initial condition) by a vector $\phi_+ \in \Phi_+$ 
and the state a system when evolving towards the past (i.e. going to a final condition) by a vector 
$\psi_- \in \Phi_-$. When considering the scattering of particles by the potential barrier, this division between initial states and final states is easily explained, but we think that it is useful for studying other phenomena, like the decay towards the equilibrium position. Since both spaces  $
\Phi _{+}$ and $\Phi _{-}$ are dense in the corresponding Hilbert space $\cal H$,
at this stage there are no empirical results that could help us  decide if the conventional representation of the state of a physical system 
by a vector in $\cal H$ is better than the representation of the same system by a vector in $\Phi_+$ or $\Phi_-$.
 Both representations  give the same empirical results, so both are, in a sense, correct.
The difference is at a mathematical level (in the topologies used, etc.) and   so we can 
use the representation we find more suitable to the description of the physical
facts.

Now, by selecting  two different mathematical structures to represent a physical system
when evolving towards the future and towards the past we have introduced the basis for a  time
asymmetry in the mathematical description of  time evolution. In fact, there are only two causes for asymmetry in nature: either the laws of nature are asymmetric or the solutions of the equations of the theory are asymmetric. E. g.: the laws governing the  weak interaction are asymmetric while the solutions of the theory are asymmetric in the case of spontaneous symmetry breaking.

Time-asymmetry is not  an exception. Thus, if we want to retain the  time-symmetric laws of nature the only reason to explain the time-asymmetry of the Universe and its subsystems is to postulate that the space of solutions is not time-symmetric, namely to use the second cause for asymmetry. So the proper way to solve the problem is simply to define a realistic time-asymmetric space of admissible physical solutions $\Phi_+$; namely to restrict the space of initial conditions. If $\hat K$ is the time inversion operator, this space will be time-asymmetric whenever $\hat K : \Phi_+ \mapsto \Phi_-$, namely time inversion changes the physically admissible solutions in $\Phi_+$ into a  space of inadmissible solutions $\Phi_-$ different to the previous one. If $\cal H$ is the usual space of solutions of the theory (Hilbert space in quantum mechanics or Liouville space in classical statistical mechanics) and we select $\Phi_+$ properly, then mathematically speaking we have introduced a Gel'fand triplet $\Phi_+ \subset {\cal H} \subset \Phi_+^\times $.

A Reichenbach branched system is perhaps the most realistic model for an irreversible Universe, i. e. a set of irreversible processes such that each one begins in an unstable state produced by another member of the system and it eventually ends in an equilibrium state \cite{Reichenbach}. This set of processes, all of them beginning in a non-equilibrium state, defines a global arrow of time in the Universe. The problem is that the whole branch system must begin in a global  unstable  initial state which has no explanation. This  unstable initial state would be the initial cosmological state of the Universe. It is qualitatively shown in ref. \cite{Davies} that the expansion of the Universe can be the agency that produces this initial unstable state. Using the method of this paper, we have found the same explanation but in a quantitative way \cite{Aquilano}, endowing the Universe with a (global) space of admissible solutions $\Phi_+$. Therefore, we think that the time-asymmetry is not given by the system itself, namely our upside-down oscillator; instead it must be connected with the global arrow of time of the Universe. In fact, our system is really a member of the Reichenbach branch system in such a way that the unstable  initial condition of our oscillator is necessarily produced by another member of the branch system. Thus, as an initial condition is given, it is  admissible only if it belongs to a (particular) space $\Phi_+$. 
 Since the Rigged Hilbert space and the conventional Hilbert Space formulations are both equally correct, we can  select the space better qualified to define   time-asymmetry.

3.- As this procedure has been applied in the past to some unstable quantum mechanical
systems (simple potential scattering problems \cite{BohmGadella} and Friedrichs'
model \cite{Prigogine}) and chaotic classical statistical systems (Renyi
maps \cite{Antoniou1} and Baker's transformation \cite{Antoniou2}), in this paper
we extend this technique to the simplest unstable system: the upside-down harmonic
oscillator. Furthermore, we show that the two sets of spaces defined in quantum mechanics
and classical statistical mechanics are connected in a simple manner. This is the first
result of this kind that we know of.

The organization of the paper is as follows. In sec. 2 we give the
principal properties of the rigged Hilbert space formulation of quantum
mechanics. We define the regular state space and the generalized state space
and we define generalized eigenfunctions. In sec. 3 we work out the
motion of a particle subject to the potential (\ref{potential}) in classical
mechanics, introducing the canonical variables we shall  use  throughout the paper. In
sec. 4 we study the same problem in the context of quantum mechanics. We
define the growing and decaying Gamow vectors and using these objects, we also define the
two RHSs that represent states evolving to the future and states evolving to
the past. In sec. 5 we undertake the same task in the context of
classical statistical mechanics. We find the ``statistical Gamow vectors'',
namely generalized density functions in state space that represent idealized
growing and decaying states. Once again, using these objects we define the two
RHSs that represent states evolving towards the past and towards the future. In
sec. 6 we show that the structures defined in sections
4 and 5 are connected by the Wigner function. 
Finally, in sec. 7 we draw our
conclusions.

\newpage

\section{The rigged Hilbert space formulation of quantum mechanics}

In the traditional (von Neumann's) formulation of quantum mechanics,  a physical state is represented by a 
vector in a Hilbert space $\cal H$ and physical magnitudes (observables) by linear selfadjoint operators acting in it.
We will deal in this paper with the unidimensional motion of a particle in a potential field, so the Hilbert 
space we should  work in is isomorphic to $L^2(\Bbb R)$. More precisely, in the position ($|q \rangle $) 
representation the particle's state is represented by a normalized wavefunction $\psi (q) \in L^2(\Bbb R)$ 
whose modulus squared gives the probability density of finding the particle in the position $q$.
Observables are then represented (in this representation) by selfadjoint operators in $L^2(\Bbb R)$. 

This formulation contains certain idealizations, since once we interpret the wavefunction as a probability 
amplitude, the only physical requirement is that it must be square integrable. But a Hilbert space is   
a complete topological space, with respect to a particular topology, namely the one obtained from its scalar product. The assumption that 
every vector in this Hilbert space represents a physically realizable state cannot, by no means, be justified by empirical 
facts, since a topology (infinite limits, continuity) has no physical meaning. It is just a mathematical 
idealization with which a theoretical physicist works in order to  formulate a theory. So  we can 
take another mathematical idealization and formulate a theory in a different mathematical environment. For instance, we can take another topological space, which is itself a subspace of the Hilbert space, and associate with the vectors 
in such a space the states of the physical system. 

This is what the rigged Hilbert space (RHS) formulation of quantum mechanics does \cite{BohmGadella}. 
To define a RHS we take a topological vector space (endowed with a nuclear topology) $\Phi$ with a 
continuous scalar product defined in it. As we shall see, when the Hilbert space is the space of square integrable 
functions, the test function space $\Phi$ is chosen as the set of functions of fast decrease or Schwarz functions ${\cal S} \equiv {\cal S}(\Bbb R)$ or some subspace of it. By completing the vector space with the topology given by the 
scalar product we get a Hilbert space $\cal H$ such that $\Phi \subset \cal H$.
If we consider the set of all antilinear functionals, continuous with respect to the nuclear topology,
we get another vector space called the dual space of $\Phi$ and denoted $\Phi^\times$. We will denote the value of the functional $F \in \Phi^\times$ on the vector $\phi \in \Phi$ by $\langle \phi | F \rangle$ and its complex conjugate by $\langle F  |\phi    \rangle$. Due to Riesz's 
lemma, the dual space of $\cal H$ is $\cal H$ itself, so we get (using the fact that the nuclear topology is 
stronger than the Hilbert space topology) the Gel'fand triplet

\begin{equation}
\Phi \subset {\cal H} \subset \Phi^\times.
\end{equation}

We associate the vectors in the space $\Phi$ with the physical states of the system in consideration, thus it is usually called the regular state space. The 
observables are associated with continuous (essentially) selfadjoint linear operators in $\Phi$.

This formulation has some  advantages over the conventional one. The first advantge is that in the traditional formulation of quantum mechanics a 
vector is a class of Lebesgue square integrable functions differing in a set of  measure zero while in the RHS 
formulation a vector is, usually, just one continuous and infinitely differentiable function.

The second advantage is that every observable is well defined, since they are continuous in $\Phi$. This means 
in particular, that even when the operator is unbounded in $\cal H$ it is well behaved in $\Phi$, which is 
contained in the domains of all  operators of interest. In the case we will be 
studying, these include the position, momentum  hamiltonian operators.

The third advantage is that every essentially selfadjoint continuous linear operator in a RHS has a complete 
set of generalized eigenvectors in $\Phi^\times$ (the generalized state space) with their eigenvalues in the spectrum of the operator, a 
result proved by Gel'fand and Maurin \cite{BohmGadella, Gelfand}.

 By a generalized eigenvector of an operator $\hat A$ in a RHS we mean a functional $|F_\lambda \rangle 
\in \Phi^\times$ such that  

\begin{equation}
\langle   A^\dagger \phi  | F_\lambda   \rangle = \lambda \langle \phi  | F_\lambda   \rangle \quad \forall \,  \phi 
\in \Phi. \label{autovgen}
\end{equation}

Since this is a generalization of the definition of an eigenvector in finite dimensional vector spaces, the 
number $\lambda$ is called a generalized eigenvalue corresponding to the eigenvector $|F_\lambda \rangle$. 
What the Gel'fand-Maurin theorem states is that given an operator, the set of generalized eigenvectors spans 
the regular vector space $\Phi$. More precisely, if $\Lambda $ is the spectrum of the aforementioned 
operator $\hat A$, then the scalar product between two vectors $\phi, \psi \in \Phi$ can be expressed as

\begin{equation}
(\phi , \psi ) = \int_\Lambda \, d\mu(\lambda) \, \langle \phi  | F_\lambda   \rangle \langle F_\lambda  | \psi   
\rangle \label{gelmau}
\end{equation}

where $\mu $ is a certain integration measure. We see then that in the RHS formulation of quantum 
mechanics, Dirac's notation is fully justified. Even though the generalized eigenvectors  do 
not belong to the Hilbert space, they are defined as antilinear functionals in $\Phi^\times$.
In the same way, the operator itself can be expressed as a linear combination of the same generalized eigenvectors, 
namely

\begin{equation}
\hat A = \int_\Lambda \, d\mu(\lambda)  \, \lambda \, | \lambda \rangle \langle \lambda | .
\end{equation}

There is one more advantage that this formulation has over the traditional one:  we can define 
generalized eigenvectors of an essentially selfadjoint operator with eigenvalues that do not belong to the 
(Hilbert space) spectrum of the operator. In the general case, such a spectrum is a closed subset of the real 
line but, as we shall show below, we can find in some cases eigenvectors with complex 
eigenvalues. This will allow us to define Gamow vectors, namely generalized eigenvectors of the 
hamiltonian operator with non zero imaginary eigenvalues. As we shall see, this fact implies that the evolution of these vectors is 
exponential, either growing or decaying, depending on the sign of the imaginary part of the eigenvalue.

These new generalized eigenvectors are useful to study the temporal evolution of the regular vectors 
representing physical states, if we define the RHS so that we can find new expansions like (\ref{gelmau}) 
containing them. We will find that in order to do so, we must define two different RHSs, that we shall call 
$\Phi_+$ and $\Phi_-$.   The first one will correspond to the representation of physical systems when they 
evolve towards the future, since its vectors will be expressed in terms of the decaying Gamow vectors,  and the second one will correspond to the representation of physical systems when they evolve towards the 
past, since they wil be expressed in terms of the growing  (or decaying towards the past) Gamow vectors.

To finish this section we will mention the different rigged Hilbert spaces with which we will work in this 
paper \cite{Gelfand}. The first one is the one constructed from the space of Schwarz class functions $\cal S$, composed of all  
infinitely differentiable functions of a real variable such that together with all their derivatives vanish at infinity faster than the inverse of any polynomial. The RHS obtained is ${\cal S} \subset L^2 (\Bbb R) \subset {\cal S }^\times$. In this case, the 
functionals in ${\cal S}^\times$ are called tempered distributions. This is the RHS  most theoretical physicists work in, since in it the position and momentum operators are continuous \cite{BohmGadella}.

The other two RHSs we shall use are constructed from  subspaces of $\cal S$, so that this last property is maintained. These subspaces are the 
space $\cal K$ of infinitely differentiable functions of a real variable with compact support and the space $\cal 
Z$ of  Fourier transforms of functions in $\cal K$, that is isomorphic to the space of entire functions of 
fast decrease. The RHSs we 
obtain are then ${\cal K} \subset L^2(\Bbb R) \subset {\cal K}^\times$ and  ${\cal Z} \subset L^2(\Bbb R) \subset {\cal Z}^\times$.

\newpage

\section{The classical upside-down oscillator}

The system we are going to study is one of the simplest unstable ones, namely
the motion of a particle in the presence of a potential  of the form (\ref{potential}).
The hamiltonian function of the system is $
H = {p^2 \over 2m} - {1 \over 2} m \omega^2 q^2 $.

As in the study of the harmonic oscillator, it is convenient, at this point, to adimensionalise
the dynamical variables. In order to do that, we must take the natural scales of  length,
momentum and time
defined through the physical parameters available: $m$, $\hbar$ and $\omega$ (we include
$\hbar$ since we will deal in the next section with the quantum case).
These scales are, respectively  $ \sqrt{ \hbar / m \omega}$ , $ \sqrt{ m \omega \hbar}$ and $ 1 /
\omega$, so we are making the transformations

\begin{equation}
q \mapsto \sqrt{m \omega \over \hbar} q  \quad , \quad p \mapsto {1 \over \sqrt{m \omega \hbar}}
p
\end{equation}

and

\begin{equation}
H \mapsto { 1 \over \hbar \omega} H \quad , \quad t \mapsto \omega t .
\end{equation}

The relation between the Hamiltonian and the adimensional $q$ and $p$ variables is

\begin{equation}
H = {1 \over 2} (p^2 - q^2) .
\end{equation}

To solve the equations of motion, it will be helpful throughout the paper to
work in another couple of canonical variables

\begin{equation}
v = {1 \over \sqrt{2 }} ( p +  q) \quad , \quad u =  {1 \over \sqrt{2 }} ( p -  q) \label{transf.can}
\end{equation}

obtained through the generating function $
F(q,v) = -{1 \over 2} v^2 + \sqrt{2 } q v  - {1 \over 2}  q^2 .$

Expressed in these new variables, the hamiltonian reads

\begin{equation}
H =  v u \label{hamilt}
\end{equation}

and thus, the equations of motion are

\begin{equation}
\dot v = {dv \over dt} = {\partial H \over \partial u} =  v \quad , \quad \dot u = {du \over dt} = -
{\partial h \over \partial v} = -  u .
\end{equation}

These equations are  uncoupled and they have as solutions for any initial condition
$ v_0 , u_0 $

\begin{equation}
v = v_0 e^{ t} \quad , \quad  u = u_0 e^{- t}. \label{soluc}
\end{equation}

In phase space, the trajectories of the particles are hyperbolic and the $v$ and $u$ axes are the
corresponding
asymptotes. The point $ v=0, u=0 $ (or $q=0,p=0$) is a point of unstable equilibrium.
If $H_0 = v_0 u_0 \ne 0 $ then the particle decays (gets away from the barrier region) both
towards the past and the future.
The directions parallel to the $v$ axis will be called ``dilating fibers'' while those parallel to the
$u$ axis will be called ``contracting fibers''.

\newpage

\section{The quantum upside-down oscillator}

\subsection{The  $| v \rangle$ and $| u\rangle$ representations }

As we have seen in the previous section, the evolution of the system is 
expressed in a simple way when we use the variables $v$ and $u$. In quantum mechanics, a canonical transformation
is  associated with a change in representation. Instead of using the representations
$| q \rangle$ and $| p \rangle$ of generalized eigenfunctions of  $\hat Q$ and $\hat P$, we will use the representations
$| v \rangle$ and $| u\rangle$ of generalized eigenfunctions of the operators $\hat V$ and $\hat U$, defined through

\begin{equation}
\hat V = {1 \over \sqrt{2}} (\hat P +  \hat Q) \qquad \hat U = {1 \over \sqrt{2}} (\hat P -  \hat Q). \label{repsvu}
\end{equation}

Since these operators are the quantum representations of canonically conjugate 
variables, they satisfy the commutation relation

\begin{equation}
[ \hat V , \hat U ] = i \, \Bbb I .\label{conmut}
\end{equation}

The spectrum of these operators is the whole real line \cite{Cohen}, and the transformations from the $|q\rangle$ representation to the $|v\rangle$ and $|u\rangle$ representations are given by

\begin{equation}
 \langle q | v \rangle =  (2 \pi^2)^{-1 / 4} e^{i ( \sqrt{2  } v q -   q^2 /2 - v^2 /2)} \label{qve}
\end{equation}

\begin{equation}
\langle q | u \rangle =  (-2 \pi^2)^{-1 / 4} e^{i ( \sqrt{2  } u q +   q^2 /2 + u^2 /2)}.\label{que}
\end{equation}

Due to (\ref{conmut}), we get the relation

\begin{equation}
\langle v | u \rangle = {1 \over \sqrt{ 2 \pi }}  e^{iuv}. \label{V-U}
\end{equation}

\subsection{Eigenfunctions of the Hamiltonian}

\subsubsection{Real eigenvalues}

The Hamiltonian of the system is

\begin{equation}
\hat H  = {1 \over 2} {\hat P}^2 - {1 \over 2} {\hat Q}^2 = { 1 \over 2 } (\hat V \hat U + \hat U \hat V)
\end{equation}

namely, the symmetric version of (\ref{hamilt}). The eigenvalue equation for the 
eigenfunctions of this Hamiltonian with eigenvalue $\epsilon$  in the $| v \rangle$ and $| v \rangle$ 
representations reads

\begin{equation}
v{d \over dv}\phi_\epsilon (v)  = (  i \epsilon - {1 \over 2}) \phi_\epsilon (v) \quad , \quad u{d \over du}\psi_\epsilon (u)  = (  -i \epsilon - {1 \over 2}) \psi_\epsilon (u). \label{eqSch}
\end{equation}

Formally, the solutions of these equations are

\begin{equation}
\phi_\epsilon (v) =\alpha v^{i \epsilon - 1/2} \quad , \quad \psi_\epsilon (u)  = \beta u^{-i \epsilon - 1/2}\label{sol.formal}
\end{equation}

but care must be taken, since these expressions are only defined for positive values of $v$ and $u$ respectively (in fact, as we have seen in sec. 2, they are not functions but distributions).

Actually, there are two linearly independent solutions of the equations (\ref{eqSch}) for each value of $\epsilon$, due to the degeneracy of the Hamiltonian. These independent solutions can be chosen as \cite{Lucho}

\begin{equation}
\langle v  | \epsilon + (v)   \rangle ={1 \over \sqrt{ 2 \pi}}  \theta (v) v^{i \epsilon - 1/2} \quad , \quad \langle v  | \epsilon - (v)  \rangle = {1 \over \sqrt{ 2 \pi}} \theta (-v) |v|^{i \epsilon - 1/2} \label{epsilon-v}
\end{equation}

or 

\begin{equation}
\langle u  | \epsilon + (u)   \rangle ={1 \over \sqrt{ 2 \pi}}  \theta (u) u^{-i \epsilon - 1/2} \quad , \quad \langle u  | \epsilon - (u)  \rangle = {1 \over \sqrt{ 2 \pi}} \theta (-u) |u|^{-i \epsilon - 1/2}. \label{epsilon-u}
\end{equation}

The generalized functionals (\ref{epsilon-v}) represent the idealized scattering out-states, representing particles leaving to the left and the right respectively, while the generalized eigenfunctions (\ref{epsilon-u}) represent the scattering in-states representing particles entering the scattering region from the left and the right respectively \cite{Balasz, Barton}. Both of these sets of solutions form complete and orthonormal sets in $\cal S$, in the sense discussed in sec. 2.

By calculating the scalar product between the in-states and the out-states, we  obtain the scattering matrix, whose coefficients are in this case of the form \cite{Lucho, Barton}

\begin{equation}
S_{\mu \nu} (\epsilon) = f_{\mu \nu}(\epsilon) \Gamma({1 \over 2} - i\epsilon) \quad \mu , \nu = + , - 
\end{equation}

where the $f_{\mu \nu}(\epsilon)$ are entire functions of $\epsilon$. 

\subsubsection{Eigenfunctions with complex eigenvalues}

As  can be seen from the  above formula, the scattering matrix when extended to the complex plane has an infinite number of imaginary poles located at  $z_n = -i   (n + 1/2) $ where $n$ is a nonnegative integer. The Gamow vectors will have as generalized eigenvalues these numbers or their complex conjugates. Instead of taking the conventional way to obtain the expressions for these Gamow vectors (namely by analytical extension of the scalar product to the complex plane \cite{Bohm}), we will take a shortcut and find them in an heuristic way.

We will consider the solutions of the eigenvalue equations (\ref{sol.formal}) when taking $\epsilon = z_n $ in the first one and $\epsilon = \overline{z_n}$ in the second one. Then we get the functionals 

\begin{equation}
\langle v | n \rangle = v^n \quad , \quad \langle u | \tilde n \rangle = {(-i)^n \over \sqrt{2 \pi} n!} u^n \label{un}
\end{equation}

where the multiplicative factors have been selected so that (\ref{biort}) below applies.
Transforming with (\ref{V-U}) we get

\begin{equation}
\langle u |  n \rangle = \sqrt{2 \pi} i^n \delta^{(n)} (u)  \quad , \quad  \langle v | \tilde n \rangle =  {(-1)^n \over n!} \delta^{(n)} (v) .\label{vn} 
\end{equation}

It is clear that these functionals are tempered distributions. By direct calculation, we can demonstrate the biorthonormality of the sets $\{ | n \rangle \}$ and $\{ | \tilde  n \rangle \}$, namely that 

\begin{equation}
\langle n' | \tilde n \rangle = \langle \tilde n | n' \rangle = \delta_{n, n'} .\label{biort}
\end{equation}

We will show now that  $| n \rangle$ is indeed a generalized eigenfunction of $\hat H$ with complex eigenvalue $z_n$, namely that $
 \langle \hat H \phi | n \rangle = z_n \langle \phi | n \rangle  \, \forall \pi \in \cal S$. 

We have

$$
\langle \hat H \phi | n \rangle = \int_{-\infty}^{+\infty} dv \, [i   (v {d \overline {\phi (v)} \over dv }+ {1 \over 2} \overline {\phi (v)})] v^n 
$$

and integrating by parts in the right hand side, we get

$$
\langle \hat H \phi | n \rangle = -i     \int_{-\infty}^{+\infty} dv \, \{ \, \overline {\phi (v)} \, (n+1) v^n - {1 \over 2} \overline {\phi (v)} \} = -i   (n + {1 \over 2}) \langle \phi | n \rangle.
$$

In a similar fashion, we can demonstrate that $| \tilde n \rangle$ is a generalized eigenfunction
of $\hat H$ with eigenvalue  $ \overline{z_n} = i   (n + 1/2)$. It is 

$$
\langle \hat H \phi | \tilde n \rangle = \int_{-\infty}^{+\infty} dv \quad [i   (v {d \overline {\phi (v)} \over dv }+ {1 \over 2} \overline {\phi (v)})] {(-1)^n \over n!} \delta^{(n)} (v)
$$

and using 
${d^n \over dv^n} (v {d \over dv}) = v {d^{(n+1)} \over dv^{(n+1)}} + n {d^n \over dv^n}$ we get 
$\langle \hat H \phi | \tilde n \rangle = \overline z_n \langle \phi | \tilde n \rangle$.

Since $| n \rangle $ and $|\tilde  n \rangle $ are generalized eigenvectors of $\hat H$, their temporal evolution is easily calculated. In fact, we get 

\begin{equation}
e^{-i \hat H t} | n \rangle  = e^{-(n+1/2)t} |n \rangle \quad , \quad
e^{-i \hat H t}| \tilde n  \rangle  = e^{(n+1/2)t} |\tilde n \rangle \label{nt}
\end{equation}

showing that the Gamow vectors are, indeed, vectors that would  represent idealized states that decay or grow in a perfectly exponential way. These functionals are more pathological (are ``less physical'') than the eigenfunctions of the Hamiltonian with real eigenvalues (\ref{epsilon-v}) or (\ref{epsilon-u}); they are clearly distributions and not regular functions and thus, cannot represent, by themselves,   physical states. It will be shown, though, that they are useful in studying the temporal evolution of the regular states.

To complete the presentation of these functionals, let us study their expression 
in the $ |q \rangle$ representation. In order to do that, we use  (\ref{que}) and get

$$
\langle q | n\rangle =  \alpha_n '   e^{i   q^2 /2} {\partial^n \over \partial u^n} ( e^{i(\sqrt{2  } u q + u^2 /2)})  |_{u=0} 
$$

where $\alpha_n ' $ is just a numerical factor. Now, using the formula $
H_n (z) = {\partial^n \over \partial \lambda^n} (e^{- \lambda^2 + 2 \lambda z})|_{\lambda = 0} $ where $H_n(z)$ is the n-th Hermite polynomial, we find that

\begin{equation}
\langle q  | n   \rangle = \alpha_n   e^{i   q^2 /2 } H_n ( e^{-i \pi /4} q). \label{q n}
\end{equation}

In an analogous way, using (\ref{qve}) we get

\begin{equation}
\langle q  | \tilde n   \rangle = \tilde{\alpha_n} e^{-i   q^2 /2 } H_n ( e^{i \pi/4}q). \label{q n monio}
\end{equation}

Restoring the variables' dimensions, we get

$$
\langle q  | n   \rangle = C_n e^{i m \omega q^2 / 2 \hbar } H_n ( \sqrt{{-i m \omega \over \hbar}}  q)
\quad , \quad 
\langle q  | \tilde n   \rangle = \tilde{C_n}  e^{-i m \omega q^2 / 2 \hbar } H_n ( \sqrt{{i m \omega \over \hbar}}  q).
$$

If we compare these generalized states with the eigenstates of the harmonic
oscillator with equal frequency and mass \cite{Merzbacher}, we can see that the former
can be obtained from the latter by means of the transformation $\omega \mapsto -i \omega$ in the case of
$|n\rangle$ and the transformation $\omega \mapsto i \omega$ in the case of $| \tilde n \rangle$. But these
transformations applied to the potential of the harmonic oscillator turn it into
the potential (\ref{potential}) and the same happens with the eigenvalues: they are transformed from discrete real eigenvalues to discrete imaginary ones.
Once again we see (now in the $| q \rangle$ representation) that these generalized eigenfunctions cannot represent physical states;
when calculating the norm squared of these functions we obtain  that they diverge 
in the limit  $|q| \mapsto \infty$  as $q^{2n}$.

\subsection{Generalized expansions and the regular function spaces $\Phi_+$ and $\Phi_-$}

We study now the use of the generalized functions (\ref{un}) in generalized expansions. Our first step towards this end is taking $\phi$ such that $\langle v  | \phi   \rangle = \phi(v) \in \cal S$. Since this wavefunction is infinitely differentiable, we can define its Taylor expansion around $v=0$

\begin{equation}
\phi(v) = \sum_{n=0}^\infty {\phi^{(n)} (0) \over n!} v^n
=\sum_{n=0}^\infty \langle v  | n   \rangle \langle \tilde  n | \phi   \rangle. \label{expansV}
\end{equation}

The equality in this formula is restricted to the values of $v$ in a interval on  the real line, namely inside  the radius of convergence of the series, so the utility of this expansion is rather limited. Only when considered in a scalar product will it proof really useful. 

Our second step is to define the  test function spaces

\begin{equation}
\Phi_+ = \{ \psi \in {\cal H} / \langle v  | \psi   \rangle  \in {\cal Z} \}  = \{ \psi \in {\cal H} /\langle u  | \psi   \rangle \in {\cal K} \} \label{PhiMAS}
\end{equation}

\begin{equation}
\Phi_- = \{ \psi \in {\cal H} /\langle v  | \psi   \rangle \in {\cal K} \}  =  \{ \psi \in {\cal H} / \langle u  | \psi   \rangle \in {\cal Z} \} \label{PhiMENOS}
\end{equation}

where $\cal K $ and $\cal Z$ are the spaces introduced in sec. 2. We can then construct two different RHSs with these spaces, namely 

\begin{equation}
\Phi_\pm \subset {\cal H} \subset \Phi_\pm^\times.
\end{equation}

 Let us take $\phi \in \Phi_+$; in this case the expansion (\ref{expansV}) is valid for all values of $v$. 
 Considering that $|n\rangle$ is a generalized eigenfunction of $\hat H$ with eigenvalue $z_n$ and the definition $(\ref{un}_1)$,  the time evolution of this vector is given by

\begin{equation}
\phi (v , t ) = \langle v | e^{-i\hat H t} \phi \rangle = \sum_{n=0}^\infty e^{-(n + 1/2)t} \langle v  | n   \rangle \langle \tilde  n | \phi   \rangle = e^{-t/2} \phi(ve^{-t}, 0). \label{phi,v,t}
\end{equation}

 Since $\Phi_+$ is a dense subspace in $\cal H$, we get that  (\ref{phi,v,t}) must be valid for any vector in $\cal H$; this can be verified by direct substitution in  Schroedinger's equation.
If we consider the Taylor expansion of a vector $\psi$ in $\Phi_- $ in the $ |u \rangle $ representation, we get that in that representation the temporal evolution is given by $\psi ( u , t) = e^{t/2}\psi(ue^t, 0)$.

These two results can be seen as the fact that the wavefunction does not change its form with time; it just suffers a change in scale. If the wavepacket is initially concentrated in some value $v_0$ (resp. $u_0$) then it will be concentrated at time $t$ in $v_0e^t$ (resp. $u_0e^{-t}$). Namely, the center of the wavepacket follows the classical trajectory found in sec. 3. This is a consequence of Ehrenfest's theorem, since the potential is quadratic in position.  Quantum effects come from the broadening of the wavefunction in the $| v \rangle$ representation and the narrowing of it in the $| u \rangle$ representation.  

Now, let us consider the scalar product between a vector $\phi_- \in \Phi_-$ and a vector  $\psi_+ \in \Phi_+$, $
(\phi_- , \psi_+ ) = \int_{-\infty}^{+\infty} \, dv \, \overline{\phi_-(v)} \psi_+(v) $.
Since $\phi_-(v) \in \cal K$, it is an infinitely differentiable function of compact support; the integration limits can then be replaced by $-a,a$ for some $a$. On the other hand, since $\psi_+(v)$ is an entire function, the radius of convergence of its Taylor expansion around $v=0$ is infinite, and so we have

\begin{equation}
(\phi_- , \psi_+ ) = \int_{-a}^a \, dv \, \overline{\phi_-(v)} \, \sum_{n=0}^\infty \langle \tilde n   | \psi_+   \rangle v^n.
\end{equation}

The convergence of the series in the interval $[-a,a]$ is uniform and, since the function  $\phi_- (v) $ is bounded in that interval, we can interchange the order of the summation and integration. Then we get

\begin{equation}
(\phi_- , \psi_+ ) = \sum_{n=0}^\infty \langle \phi_-  |  n  \rangle \langle \tilde n  |  \psi_+  \rangle. \label{menosmas}
\end{equation}

We get a  second expansion of this kind by taking the complex conjugate of this expression:

\begin{equation}
(\psi_+ , \phi_-) =  \sum_{n=0}^\infty \langle \psi_+  | \tilde n  \rangle \langle  n  |  \phi_-  \rangle. \label{masmenos}
\end{equation}

This shows that  a vector $\phi_+ \in \Phi_+$ can be expanded  (when acting as a functional in $\Phi_-^\times $) as $ |\phi_+ \rangle = \sum | n \rangle \langle \tilde n | \phi_+ \rangle$
 and a vector $\phi_- \in \Phi_-$ can be expanded as  $ |\phi_- \rangle = \sum | \tilde n \rangle \langle  n | \phi_- \rangle$.

Let us now turn to the physical meaning of these RHSs. As  we have seen,  if $\phi_+ \in \Phi_+$ the expansion (\ref{expansV}) is valid for all values of $v$. This means that we can represent a state as an infinite sum of decaying states. Then, if we want to study the decay of a physical system we can represent its initial condition by a vector $\phi_+ \in \Phi_+$. If on the other hand, we want to study the creation of a physical system then by representing the final condition with a vector $\psi_- \in \Phi_-$ we can think of this process as the growth of a linear combination of the exponentially growing states $|\tilde n \rangle$.

In an experiment, we control the initial condition, wich evolves towards the future, and then compare it with a final condition by means of a measurement; thus, the quantities of interest are of the form $( \psi_- , \phi_+(t)) = \langle \psi_- | e^{-i \hat H t }| \phi_+\rangle$, wich can be expanded as in (\ref{menosmas}) 

\begin{equation}
\langle \psi_- | e^{-i \hat H t }| \phi_+\rangle = \sum_{n=0}^\infty e^{-(n+1/2)t} \langle \phi_-  |  n  \rangle \langle \tilde n  |  \psi_+  \rangle. \label{menosmast}
\end{equation}

It is evident from this formula that we can think that the initial condition is fixed and the final condition evolves to the past. The series in the formula converges always (for all values of $t$) as we have seen, but it only has physical meaning for positive values of $t$. For long times, only the first term of the series makes a contribution and then we can take $( \psi_- , \phi_+(t)) = e^{-t/2} \langle \phi_-  | 0  \rangle \langle \tilde 0  |  \psi_+  \rangle $ showing that in this regime the decay is effectively exponential, unless one of the coefficients  $\langle \phi_-  | 0  \rangle $ or $ \langle \tilde 0  |  \psi_+  \rangle$ vanishes. We recover thus the lifetime already known from reference \cite{Barton}. We remind the reader  that equations (\ref{menosmas}) and (\ref{menosmast}) are exact, there are no approximations in the series. In our system, unlike the one considered in \cite{BohmGadella} there is no background term; all the details of the evolution are found in the series expansions.

Finally, let us consider time reversal in our formulation. From the interpretation we have given to the spaces $\Phi_+ $ and $\Phi_-$, it seems that the time reversal operator should transform one of the spaces into the other, since it changes initial conditions into final conditions. This we shall show now.

Wigner's time reversal operator  is defined in the $| q \rangle$ representation as the conjugation operator, namely

\begin{equation}
\hat K : |\phi \rangle  \mapsto  | \phi ' \rangle \, {\rm where} \, 
 \phi ' (q) =\overline{ \phi (q)}.
\end{equation}

Let us take $\phi \in \Phi_+$. Then we have

\begin{equation}
\langle v | \phi \rangle = \phi (v) ={ 2^{1/4} \over \sqrt{2 \pi} } \int_{-\infty}^{+\infty} dq e^{i(-\sqrt{2 } v q +  q^2 /2 + v^2 /2)}  \phi (q) \label{K1}
\end{equation}

\begin{equation}
 \varphi '(u) =\langle u | \phi ' \rangle ={ 2^{1/4} \over \sqrt{2 \pi i} } \int_{-\infty}^{+\infty} dq e^{-i(\sqrt{2 } v q -  q^2 /2 - v^2 /2)}  \overline{\phi(q)}. \label{K2}
\end{equation}

By comparing (\ref{K1}) and (\ref{K2}), we get
 
\begin{equation}
 \varphi ' (u) = e^{-i \pi /4} [\overline{\phi (v)}] \Big |_{v= -u}
\end{equation}

and considering (\ref{PhiMAS}) and (\ref{PhiMENOS}), $ \phi ' \in \Phi_-$ , or

\begin{equation}
\hat K : \Phi_+ \longmapsto \Phi_-.
\end{equation}

The inverse relation can be proved in a similar way.

\newpage

\section{The upside-down oscillator in classical statistical mechanics}

\subsection{Generalized eigenfunctions in classical statistical mechanics}

In classical statistical mechanics, a state of a physical system is represented by a density function $\rho (v,u)$ that gives the probability density to find it in a given region in phase space. This density function must satisfy certain requirements to have a physical meaning. First of all, it must be a positive function, since probabilities must be positive. Secondly, it must be integrable over the entire phase space. Finally, we  ask the further condition that it be square integrable, since we want to calculate mean values of functions that will be themselves square integrable functions. In this case, we work in a Hilbert space. Just like we did in quantum  mechanics, we will formulate the classical statistical theory in a rigged Hilbert space rather than in a Hilbert space, thereby introducing a time asymmetry in the representation of physical states. This method has been applied before to simple chaotic problems  \cite{Antoniou1, Antoniou2}.

Let us consider the equation that gives the temporal evolution of the density functions, namely the Liouville equation

\begin{equation}
i{\partial \rho \over \partial t} = \hat L \rho \label{Louiville}
\end{equation}

where the Liouvillian operator is defined by 

\begin{equation}
\hat  L \rho = i \{ H , \rho \}_{P.B.} = i( \frac{ \partial H}{ \partial v} \frac
 { \partial \rho}{ \partial u} - \frac{ \partial H}{ \partial u}
 \frac{ \partial \rho}{ \partial v}) .\label{Louivilliano}
\end{equation}

(\ref{Louiville}) is formally equivalent to the  Schroedinger equation, so the solutions to this equation can be found in a similar way. The Liouvillian is an (essentially) selfadjoint operator in the Hilbert space of square integrable functions of two real variables $L^2({\Bbb R}^2)$. We will find the generalized  eigenfunctions of this operator that play a similar role as (\ref{vn}) did in sec. 4.

The Liouvillian operator is, in our case, $
\hat L \rho  = i  ( u \frac{\partial \rho}
  {\partial u} - v \frac{\partial \rho}{ \partial v}) 
$
Taking $\nu$ as the eigenvalue of $\hat L$ and proposing as a solution the product of a function of $v$ by a function of $u$ ($\rho = V(v) U (u)$), we get

\begin{equation}
u U^{-1} U^\prime - v V^{-1} V^\prime = -i  \nu.
\end{equation}

It can be seen then, that both terms on the left hand side of this equation must be constants, that we shall call $m$ and $n$, respectively:

\begin{equation}
u U^{-1} U^\prime = m, \qquad v V^{-1} V^\prime = n .\label{umvn}
\end{equation}

We see that the  equations for $U$ and $V$ are the same, so they will have similar solutions.
A solution obtained by direct integration is $
V = A v^n ,\quad U = B u^m$ where $A$ and $B$ are arbitrary constants. We get then that $\rho = C v^n u^m$ is a generalized eigenfunction of the Liouvillian operator with eigenvalue  $\nu = i (m-n)$. As we did in sec. 4, let see what happens when we take $m$ and $n$ to be nonnegative integers; in this case the eigenvalue is imaginary. We will denote these eigenfunctions as

\begin{equation}
| m , n \rangle = {1 \over n! m!} v^n u^m \quad , \nu = i (m-n) .\label{MN}
\end{equation}

They are merely polynomials in $v$ and $u$. Once again, we find that this functions have no direct physical meaning; in fact, they do not satisfy any of the properties we asked for the  physical density functions.

Using the relation 

$$
v \delta^{(n+1)}(v) = -(n+1) \delta^{(n)}(v)
$$

we can see that  $V = \delta^{(n')} (v)$ is another solution to $(\ref{umvn}_2)$  if we take $n=-(n'+1)$. We find then  the following  generalized eigenfunctions of the Liouvillian operator with their corresponding eigenvalues

\begin{equation}
| \tilde m , \tilde n \rangle = (-1)^{m+n} \delta^{(m)} (u) \delta^{(n)}(v) \quad ,\nu = -i(m-n) \label{TMTN}
\end{equation}

\begin{equation}
|  m , \tilde n \rangle = {(-1)^n \over m!} \delta^{(n)} (v) u^n \quad , \nu = i (m+n+1) \label{MTN}
\end{equation}

\begin{equation}
| \tilde  m , n \rangle = {(-1)^m \over n!} \delta^{(m)} (u) v^n \quad , \nu = -i(m+n+1) .\label{TMN}
\end{equation}

These generalized eigenfunctions are biorthonormal by pairs, namely

\begin{equation}
\langle m, n | \tilde m' \tilde n' \rangle = \langle m , \tilde n | \tilde m' , n' \rangle = \delta_{m,m'} \delta_{n,n'} .
\end{equation}

All this functionals are tempered distributions, acting on the space ${\cal S}^2 \equiv {\cal S} ({\Bbb R}^2)$.

 \subsection{Rigged Hilbert Spaces and time asymmetry in Classical Statistical Mechanics}

With the generalized eigenfunctions just found we can construct four different function spaces. Nevertheless, to find time asymmetry we must work with (\ref{MTN}) and (\ref{TMN}). It has to be that way, since we must treat in a different way the contracting and dilating fibers, which in this case are the $u$ and $v$ variables.

On the other hand, let us consider the temporal evolution of these generalized eigenfunctions of the Liouvillian. We get

\begin{equation}
| m , \tilde n (t) \rangle  = e^{(m+n+1)t}  | m , \tilde n \rangle \label{mTnt}
\quad , \quad 
| \tilde m , n (t) \rangle = e^{-(m+n+1)t} |\tilde  m ,  n \rangle \label{Tmnt}
\end{equation}

namely, the first ones correspond to idealized states that grow exponentially towards the future while the second ones represent idealized states that decay exponentially towards the future. These are the ``statistical Gamow vectors'' that we wanted to find. In a completely analogous way to what we did in sec. 4, using (\ref{Tmnt}) we find that the solution to the Liouville equation is $\rho ( v , u, t) = \rho (v e^{-t} , u e^t, 0)$.

We define, as in sec. 4, the space of vectors that represent physical states when studying their evolution to the past and to the future, by

\begin{equation}
\Psi_+  = \{ \rho (v,u) \in L^2 / \rho \in {\cal Z}(v) \otimes {\cal K}(u) \}
\end{equation}

\begin{equation}
\Psi_-  = \{ \rho (v,u) \in L^2 / \rho \in {\cal K}(v) \otimes {\cal Z}(u) \}.
\end{equation}

Let us now consider the scalar product in $L^2({\Bbb R}^2)$

$$ 
(\rho , \rho') = \int_{-\infty}^{+\infty} dv \int_{-\infty}^{+\infty}du \, \overline{\rho(v,u)} \rho'(v,u) 
$$

where the conjugation has no effect since the functions are real. Given $\rho_+ \in \Psi_+$, we get

$$
\rho_+ (v,u) = \sum_{n=0}^\infty {v^n \over n!} {\partial^n \rho_+ \over \partial v^n} (0,u) 
$$

where the coefficients of the series are infinitely differentiable functions of $u$, that vanish outside  a given interval $[-a,a]$ of the real line. Similarly, if $ \rho_- \in \Psi_-$ then

$$
\rho_- (v,u)  = \sum_{m=0}^\infty {u^m \over m!} {\partial^m \rho_- \over \partial u^m} (v,0) 
$$

where now the coefficients of the series are infinitely differentiable functions of $v$ that vanish outside a different interval $[-b,b]$ of the real line.

Working as we did in sec. 4, we find

$$
(\rho_- , \rho_+ ) = \sum_{m,n=0}^\infty {1 \over n! m!} \big ( \int_{-\infty}^{+\infty}du \,  {\partial^n \rho_+ \over \partial v^n} (0,u)  u^m \big ) \big ( \int_{-\infty}^{+\infty}dv \,  {\partial^m \rho_- \over \partial u^m} (v,0)  v^n \big )
$$

\begin{equation}
(\rho_- , \rho_+) = \sum_{m,n=0}^\infty \langle \rho_- | \tilde m, n \rangle \langle m , \tilde n | \rho_+ \rangle. \label{1}
\end{equation}

In a restricted sense (when used in an expression between a function  $\rho_+ \in \Psi_+$ and a function $ \rho_- \in \Psi_-$) we get

\begin{equation}
{\Bbb I} =  \sum_{m,n=0}^\infty | \tilde m, n \rangle \langle m , \tilde n |.
\end{equation}

Considering the complex conjugate of (\ref{1}) we get

\begin{equation}
{\Bbb I} =  \sum_{m,n=0}^\infty |  m, \tilde n \rangle \langle \tilde m , n |.
\end{equation}

Just as in the quantum mechanical case, the justification for the distinction between the mathematical representations of a physical state of the particle, when studying its future evolution and its past evolution, comes from the fact that in the former case we want the expansion in terms of decaying states $(\ref{Tmnt}_2)$ to be valid and in the latter case we want the expansion in terms of the growing states $(\ref{mTnt}_1)$ to be valid.

Let us now face  the problem of time reversal in classical statistical mechanics. Once again we shall find that the time reversed version of a function in $\Psi_-$ belongs in $\Psi_+$ and vice versa. In this case it is easier to see, since time reversal in classical mechanics is just the transformation

$$
q \mapsto q \, , \, p \mapsto -p \, , \, t \mapsto -t
$$

and then the transformation for the variables $v$ ad $u$ are

\begin{equation}
v \mapsto - u \, , \, u \mapsto -v
\end{equation}

and so, aside for sign change, one of the variables is transformed into the other. Looking at the definitions of the spaces $\Psi_\pm$ we get that $ \Psi_\pm \mapsto \Psi_\mp$.

\newpage

\section{Connection between the classical and quantum cases}

To complete our study on the introduction of a time asymmetry in an  upside-down  simple harmonic oscillator system, we will show how the structures defined in  quantum mechanics (sec. 4) and in classical statistical mechanics (sec. 5) are connected. This connection is given by the Wigner function. This function is defined from the density matrix $\hat \rho$ representing a quantum  system by the formula

\begin{equation}
F_\rho(q,p) = (2 \pi)^{-1} \int^{+\infty}_{-\infty} dy \langle q-y/2  | \hat {\rho}
 | q+y/2 \rangle e^{ipy}. \label{Wig}
\end{equation}

It gives an idea of the probability density in  classical phase space if the  system is represented by $\hat \rho$ \cite{Wigner}. In our case, we are considering pure states, so the density matrixes are $\hat \rho = | \psi \rangle \langle \psi |$ and then (\ref{Wig}) reads

\begin{equation}
F_\psi (q,p) =(2 \pi)^{-1} \int_{-\infty}^{+\infty}  \, dy \, \psi(q - y/2)   \overline{\psi(q + y/2)} e^{ipy} .
\end{equation}

Since we have been working mostly in the $| v \rangle$ and $| u \rangle$ representation, it will be useful to express the Wigner function in this variables. It can be shown \cite{Balasz} that the Wigner function is invariant under linear canonical transformations, then we can take

\begin{equation}
F_\psi (v,u) = \pi^{-1} \int_{-\infty}^{+\infty}  \, dy \, \psi(2v - y)   \overline{\psi( y)} e^{iu(2y-2v)}  .\label{Wig2}
\end{equation}

Let us consider that  $|\psi \rangle \in \Phi_+$ and then  $\psi(v) \in \cal K$. If the support of $\psi(v)$ is $[-a,a]$, then the integration will be performed in this interval. We can see that Wigner function is infinitely derivable in its two variables, since we can make the derivation inside the integration sign.

Now, if we fix the $u$ variable and look at the dependence on the $v$ variable, then we can see  that
 it vanishes if $|v| > a$. In fact, given that $
|2v - y| \ge |2v| - |y| > a \quad \forall y \in [-a,a] 
$,  there is no interval in the real line that contributes with a non vanishing term to the integral. 

On the other hand, let us fix the $v$ variable. We see then that the Wigner function as a function of $u$ is the Fourier transform of an infinitely differentiable function of bounded support, and thus, is a function in $\cal Z$. We found then the relation

\begin{equation}
|\psi \rangle \in \Phi_+ \longmapsto F_\psi (v,u) \in {\cal K}(v) \otimes {\cal Z}(u) = \Psi_+.
\end{equation}

In a similar way, we can get the relation connecting the two spaces that describe the physical systems when they evolve to the past

\begin{equation}
|\psi \rangle \in \Phi_- \longmapsto F_\psi (v,u) \in {\cal Z}(v) \otimes {\cal K}(u) = \Psi_-.
\end{equation}

This result is not unexpected, since in our case the potential is quadratic in the position. Under this circumstances, the equation that rules the temporal evolution of the Wigner function coincides with the Liouville equation \cite{Wigner}, so there must be a close relationship between the structures found in the classical statistical case and the density function we get from the quantum case.

\newpage

\section{Conclusions}

We found that in order to give meaning to Gamow vectors, two different rigged Hilbert 
spaces must be defined: one to represent states evolving towards the future from
 initial conditions and one to represent states evolving towards the past from final conditions.
 This distinction between initial conditions and final conditions must be thought 
 of as an implementation of the global arrow of time of the Universe and not as 
 a manifestation of an intrinsic irreversibility pertaining to the system. Nevertheless, the instability
 is the key to the implementation of the time asymmetry by causing the existence of 
 Gamow vectors.

 By using this time-asymmetric formulation of the problem, we found that new 
 ``generalized expansions'' can be used to make calculations in an easy way. For example, we found the
 solution of the Schroedinger and Liouville equations by means of the Gamow vectors. 
The fact that we restrict the space of vectors representing physical states has
no empirical consequences since the test function spaces we work in  are dense in
the corresponding Hilbert space the conventional theory is formulated
in. 

Even though our model  has some characteristics that make it unpleasant, like a potential not
bounded from below, we think that these characteristics are not so hard to be avoided. 
 For instance, if we study the motion in the region around a maximum of  a twice  differentiable potential function then the approximation by a quadratic
 potential is valid. Particularly, this happens when the decay from the unstable equlibrium position is studied.

As we have stated in the introduction, this same method has been applied in the past to some models in quantum mechanics and classical statistical mechanics. In the future we will try to generalize our results to more general models in both contexts and study the conceptual implications of the time-asymmetric formulation of the theory.

\end{document}